\documentstyle[preprint,aps]{revtex}
\begin{document} 
\tightenlines
\draft

\title{ $\eta$-Nucleon Scattering Length and Effective
Range uncertainties.}
\author{A.M.~Green\thanks{email: anthony.green@helsinki.fi}}
\address{Helsinki Institute 
of Physics, P.O. Box 64, FIN--00014 University of Helsinki, Finland}
\author{S.~Wycech\thanks{email:  wycech@fuw.edu.pl}}
\address{Soltan Institute for Nuclear Studies, Warsaw, Poland}
\date{\today}
\maketitle
~\begin{abstract}  
The coupled $\eta N$ , $\pi N$, $\gamma N$, $\pi \pi N$ system is described by a $K$-matrix  
method. The parameters in this model are adjusted to get an optimal
fit to $\pi  N\rightarrow \pi  N$, $\pi  N\rightarrow \eta N$,
$\gamma N\rightarrow \pi N$ and
$\gamma N\rightarrow \eta N$ data in an energy range of about 100 MeV or
 so each side of the $\eta$-threshold.  Compared with our earlier
analysis, we now utilize recent Crystal Ball data.
However, the outcome confirms our previous result that the 
 $\eta$-nucleon scattering length ($a$) is  large with a value of
 0.91(6)+$i$0.27(2) fm. \\

\end{abstract} 


\maketitle
\section{Introduction}
\label{sec1}

The value of the  $\eta$-nucleon scattering length ($a$) is still uncertain,
but with everyone agreeing that it is indeed attractive {\it i.e.} $a>0$.
In the literature estimates can be found ranging from  
  Re$a$  of   0.3 $ \pm $  0.05 fm  
upto  about \mbox{1.0$\pm 0.1$ fm}
 -- a selection being given in Table~\ref{table1}.

The main interest in $a$ lies in the fact that, 
if the $\eta$-nucleon  scattering amplitude in the 
threshold region  is sufficiently attractive, then
$\eta$-nuclear quasi-bound states may be possible. These were first
suggested about 20  years ago \cite{hai86,li}. 
Since then many articles
have appeared on this subject studying different reactions in which
such quasi-bound states could manifest themselves. 
A bound state was  indeed
predicted in the simplest case {\it i.e.} the  
$\eta$-deuteron system \cite{ued}.
On the other hand, experimental studies of the  
$ pn \rightarrow d \eta$ cross section \cite{cal}, 
do not indicate any such bound $d \eta $ system \cite{Speth,GW01,PEN02}.
In the heavier $^{3}He$  nucleus,  the    $ pd \rightarrow \  ^{3}He \eta$ reaction 
 suggested  the likelihood of such a state    \cite{wil93,may96}. 
However,  the first experimental attempt to discover $\eta$-states in 
larger  nuclei gave a negative conclusion  \cite{chrien}. Another 
experiment is now being undertaken to check this result \cite{hayano}.

Unfortunately, in the absence of $\eta$-beams,  the formation  reactions are
also the only source of experimental information about the $\eta N$
scattering length. Therefore, in any discussion, it is  important that 
as many reactions as possible are treated simultaneously.
 Otherwise
success with one reaction may be completely nullified by failure with
another.

With this in mind, in Ref.\cite{gw97}, the present authors 
carried out a
simultaneous $K$-matrix fit to the $\pi N\rightarrow\eta N$ cross
sections reviewed by Nefkens \cite{Nefkens} and the $\gamma p\rightarrow\eta p$
data of Krusche  {\it et al.} \cite{Krusche}. In addition, the fit included   
$\pi N$ amplitudes of Arndt  {\it et al.} \cite{Arndt}, since the $\pi N$ and 
$\eta N$ channels are so strongly coupled. Using the notation for the 
elastic $T$-matrix:
\begin{equation} 
\label{0} 
T_{\eta \eta}^{-1}+iq_{\eta}=1/a+\frac{r_0}{2} q^2_{\eta}+s q^4_{\eta}
\end{equation}
--- with $ q_{\eta}$ being the momentum in the $\eta N$ center-of-mass ---
    resulted in the parameters

\noindent$a$(fm) = 0.75(4)+$i$0.27(3), $r_0$(fm) = --1.50(13)--$i$0.24(4)  and
$s$(fm$^3$) = --0.10(2)--$i$0.01(1).

\noindent A later paper by the same  authors \cite{gw99} developed this 
formalism to  include four  explicit channels :  
$ \gamma N , \pi N , \pi \pi  N $ and $ \eta N $
 and new experimental data from GRAAL \cite{ren02}. 
As indicated in Table~\ref{table1} the 
scattering length was increased  to  
$a$(fm)  = 0.87(4)+$i$0.27(2)   largely as a result of the new photoproduction 
data taken at higher energies. 
These large scattering lengths arise  as an interplay 
of the attraction induced by the $N(1535)$ state  and 
an additional attractive interaction of unknown origin. These 
generate the $\eta N $ threshold \mbox{effect} at the center-of-mass energy 1487.0
MeV. 
The  threshold enhancement as  seen in Fig.~1
  is rather  
narrow on this energy scale. Details of this figure will be discussed
later.
This enhancement may  dominate  the physics of 
few-nucleon-$\eta$  systems, but it is not  necessarily the case.
The energy region that really matters 
covers a range of negative energies, which begins at  the   
nucleon separation energy   of about 10 MeV 
and extends down by  approximately a further  $10$ MeV  due to  the 
recoil energy  of  $N\eta$ pairs - making the center-of-mass energy of
about 1460 MeV more relevant than the actual threshold energy of 1487.0
MeV.  This point may be essential to understand a  discrepancy 
between phenomenological  $ ^{3}He \ \eta$ scattering lengths  
 and four-body  calculations based on a plausible  $N \eta$
scattering amplitude. 
The phenomenological  analysis --- based upon  elastic
\mbox{$pd\rightarrow $ $ ^3$$He$$\eta$}  \cite{may96} 
and inelastic $pd\rightarrow $  $^3$$He$$\pi$ \cite{MAG03}
reactions --- produces $ A( ^{3}He \ \eta) =  4.24(29) + i 0.72(81)$ fm, 
\cite{GW03}. 
On a smaller data set a similar  
$ A( ^{3}He \ \eta)= \mid 4.3(3)\mid + i 0.5(5)$ fm
is obtained in Ref. \cite{SIB03}.
On the other hand,  a refined four-body calculation, \cite{FIX03}, 
based on a $ N \eta $ scattering  
amplitude dominated  by the $N(1535)$ and fixed to 
$a(fm)$ = 0.50+$i$0.32 fm  finds   
$ A( ^{3}He \ \eta) = 1.82 + i2.75$ fm.  
This difference in Im$ A( ^{3}He \ \eta)$ indicates an  uncertainty in the 
absorbtive, subthreshold,   $N \eta$  scattering amplitude. 
Possible effects of the  subthreshold region are also exemplified 
by calculations of the  $ d \eta $ amplitude in Ref. \cite{Shev01}.
Recent calculations of $ d \eta $ final state scattering performed  
in Ref. \cite{PEN02} 
indicate  restrictions,  0.42 $<$ Re$a ({\rm fm}) \ <  0.72 $ imposed  
by the experimental data.
All these few-body calculations are obtained with the help of separable 
model extensions into the subthreshold region. In view of the complicated 
multichannel coupling to the $N(1535)$ resonance, {\it e.g.} the effect of
the $N \eta$ threshold  and also interference of the resonant and 
potential scattering, 
the subthreshold extrapolation from the scattering length value  may be 
quite uncertain. 
Here, to provide amplitudes in this region one first uses the on-shell 
$K$-matrix approach. Next,  the effective range expansion  
of on-shell  amplitudes is calculated and off-shell 
amplitudes are generated by a simple separable  model.

We recapitulate briefly the phenomenological  model used to describe 
the $S$-wave interactions.  
The  $K$-matrix is   assumed to be of the form 
\begin{equation}
\label{1}
 K_{\alpha\beta }= B_{\alpha,\beta  }+  
\Sigma_i  \frac{\sqrt{\gamma_{\alpha}(i) \gamma_{\beta }(i)}}{E_i-E},
\end{equation}
where the sum $i= 0,1$ extends over the two  states  $N(1535)$ and $N(1650)$.
The $E_{i}$ are the positions of poles that in a ``conventional" model
should be near the energies of the $S$-wave $\pi N$ resonances $N(1535)$
and $N(1650)$. The $\gamma_{\alpha}(0,1)$ are channel coupling
parameters that are related to the widths of these resonances. Again
these widths are thought to be more or less known, when data is analysed by a 
conventional model. However, less conventional
models can lead to widths that are quite different \cite{gw98}. 
Finally the $B_{\alpha,\beta}$ are assumed, at first, to be  
energy independent background terms and are purely 
phenomenological.
However, later we shall relax this by putting a theoretically motivated 
energy dependence into $B_{\pi,\eta}$. The list of free parameters 
contains  2 resonant energies, 5 couplings to the resonant states 
 $\gamma_{\pi}(0,1),\gamma_{\eta}(0),\gamma_{\gamma}(0)$
and $\gamma_{3}(0)$. Here  $ \gamma_{3}(0)$  describes  small effects  
of the  three body $ \pi \pi  N $ channel.  
The additional 4  background parameters are $ B_{\eta,\eta}, B_{\pi, \eta}, B_{\gamma\eta}, B_{\gamma \pi}$. 
In comparison to the singular terms in Eq.~(\ref{1}), these  background terms
turn out to be very small 
with the exception of $ B_{\eta,\eta}$, which generates a  sizeable 
contribution to the  value of the $\eta N$
scattering length and the values of the scattering amplitudes at 
negative energies.  

In this note,  the calculation is extended to include the new 
Brookhaven Crystal Ball data for the
$\pi N\rightarrow\eta N$ cross section measured  close to the $\eta N$ 
threshold  \cite{koz03a,koz03b,koz03c}. These now replace the ones from the review by
Nefkens \cite{Nefkens}, used in Refs.~\cite{gw97,gw99}, and  also those  
earlier Brookhaven data from Refs.~\cite{Igor,Tom}. The latter were 
used by the present authors in Ref.~\cite{gwunc} and gave rise to
apparent massive uncertainties in the $\eta N$ effective range parameters. 
 In comparison to older measurements the new data offer 
better statistics, 
 careful estimates  of the systematic uncertainties and specified  
errors in  pion beam momenta. 
An extension  in the  model  involves  an  additional parameter $\gamma_{\eta}(1) $ 
that couples  the  $\eta N$ channel to  $N(1650)$.
Another   improvement  involves the  background terms. 
These  are  due to cross-symmetric 
amplitudes 
\footnote{We wish to thank Professor U. Mosel for indicating this point.}
and potential  interactions related to heavy meson exchanges and other  
unknown mechanisms.
On the phenomenological level  there is no need to specify these
explicitly. However, such  mechanisms 
may induce some energy dependence to  be accounted for. 
One expects the related effective  range to result from the nearest singularities in the
$u, t$ channels.
In the region of interest,  extending  from   $100$ MeV below up to  
$150$ MeV above 
the the $\eta N$  threshold, it is  the $a_0$    meson  exchange  and  
$u$-channel nucleon pole that contribute to the effective range 
in the $K$-matrix. In view of the  smallness of the  $N \eta N$ coupling
constant these 
mechanisms are expected to 
contribute an effective range mainly in the crossing $\pi ,\eta $ transitions. 
The  values of this range  follow from 
the Born amplitudes describing the nearest singularities 
\begin{equation}
\label{2}
 V_{\pi ,\eta } =  \frac{ C}{ 1 +  Q^2 / \Lambda(E)^2} \ ,
\end{equation}
where $Q$ is a momentum transfer and $C$ a constant. Projected onto 
$S$-waves such terms 
lead to logarithmic singularities for unphysical amplitudes. 
In our effective range expansion these unphysical singularities 
 are represented by a pole.
In the case of  $a_0 $ meson  exchange  $\Lambda = m_{a_0}  \approx 5$ fm$^{-1}$
and   the nearest $u$-channel singularity,  the nucleon pole, 
generates   $\Lambda \approx 3.5$ fm$^{-1}$. This sets the scale
of effective range values. In terms of the  parameter $R$, to be introduced 
 later in Eq.~(\ref{k8910}),  we obtain 
$ R_{\pi ,\eta } \approx   1/ \Lambda  \approx 0.2$ fm    
as an order of magnitude estimate.

\section{The choice of data and Results}
\label{sec2}
Since the procedures followed in this current work are very similar to
those found in Ref.~\cite{gw99}, the interested reader should refer to
that article for details concerning the fitting and error extraction.
Also much of the data is the same as we used in previous analyses, namely, 
$\pi N\rightarrow \pi  N$ providing 23 data points in the center-of-mass
energy range
\mbox{$1369.2 \leq E_{\rm cm} \leq 1705.0$} MeV \cite{Arndt},
$\gamma N\rightarrow \pi N$ 16 data points with 
$1352.0 \leq E_{\rm cm} \leq 1546.3$ MeV \cite{Arndt} and 
$\gamma N\rightarrow \eta N$ 38 data points with 
$1487.0 \leq E_{\rm cm} \leq 1523.8$ MeV \cite{ren02}. The new
ingredient is the recent Brookhaven Crystal Ball data in 
Refs.~\cite{koz03a,koz03b,koz03c}. These data are from two targets --- 
hydrogen~\cite{koz03a,koz03b} and Polyethylene~\cite{koz03c}. Both sets
are consistent with each other, but  here we use the latter, since there a
more detailed discussion is made concerning  uncertainties in the 
pion beam momentum. However, it should be added that the question of 
how the spread in the pion beam momentum affects the final value of the 
cross section is not yet fully resolved to the satisfaction of all the
authors in Refs.\cite{koz03a,koz03b,koz03c}. Fortunately, as we
shall see later, such refinements are not expected to change any of the 
 conclusions arrived at in this paper. Therefore,
for $\pi N\rightarrow\eta N$ we use the 5 data points in the energy range
$1488.5 \leq E_{\rm cm} \leq 1523.3$ MeV in Ref.~\cite{koz03c}. These are
shown in Table~\ref{CBdata}. 
We give both the uncorrected $\pi$-beam momentum and that
containing a correction suggested by the authors. Since there are
uncertainties in arriving at such corrections, we check that our final
results are not dependent on the precise values of the beam momenta. 
The errors quoted on the cross sections are the incoherent combination
of the statistical and systematic errors given in Ref.~\cite{koz03c}.

Here we present  several sets of results  in Table~\ref{results}.
The first set (A) fits all of the data given in the previous section using
an energy independent form for describing the $\pi  N\rightarrow \eta N$ 
channel {\it i.e.} $R_{\pi,\eta}=0$ in  Eq.~(\ref{k7}) of the Appendix. 
The second set (B) fits the same data as  A but without the pion beam
momentum correction in the $\pi N\rightarrow\eta N$ data from
Ref.~\protect\cite{koz03c} given in Table~\ref{CBdata}.
It is seen that the two sets have essentially the same  effective range 
parameters. Furthermore, as seen in Fig.~1, the same is
true for the $\eta N\rightarrow \eta N$ amplitude away from the
threshold. This means that, with the present data, the precise values of
the $\pi$-beam momentum are not important for extracting this amplitude.
However, the two sets give significantly different fits to the 
actual $\pi N\rightarrow\eta N$ data as seen in Fig.~2.
The fit to the beam corrected data is much better with a $\chi^2$/dp for
these 5 data points being 
about 0.2 compared with the uncorrected value of about 1.8. However, the
corresponding overall $\chi^2$/dp for the complete set of 121 data points 
are 1.01 versus 1.08.

As a further study of the importance of the new $\pi  N\rightarrow \eta N$
cross sections of Ref.~\cite{koz03c}, in set C we omit  
the Crystal Barrel data from set A. We see that, within the errors, the three
complex parameters $a, \ r_0$ and $s$ are unchanged. Also, as seen in 
Fig.~1 
the $\eta N$ amplitude is essentially unchanged.
Therefore, one's first impression is that the new data of  
Refs.~\cite{koz03a,koz03b,koz03c} adds
little to our understanding of the $\eta N\rightarrow\eta N$ amplitude.
However, this is not true. Firstly, the error bars in set C are much
larger than those in A {\it i.e.} the new data leads to tighter bounds on the
$\eta N$ amplitude. Secondly, the new data stabilizes the fit and makes it, as
far as we can see, unique. We say this because the omission of the new
data can lead to, at least, one other solution for the $\eta N\rightarrow\eta N$
amplitude with a $\chi^2$/dof=1.054 only slightly larger than the 1.045
of set C. This is shown  as set D and is quite different to the earlier
sets, but is rather similar to those extracted in Ref.~\cite{mos98} 
using a Lagrangian model --- see Table~\ref{table1}. However,
it should be added that in Ref.~\cite{mos98} the authors basically use
the  $\pi N\rightarrow\eta N$ data in Ref.~\cite{Nefkens}, which reviews
measurements made in the 1970's. These authors 
themselves point out that the data they are forced to use is poor.
Furthermore, the new
data is much nearer the $\eta N$ threshold --- the energy range that is
most important for extracting the $\eta N$ scattering parameters.
The uncertainties that can arise when using poor $\pi N\rightarrow\eta N$
data was noted earlier by the present authors in Ref.~\cite{gwunc}.
There we used some preliminary  Brookhaven non-Crystal Ball data from
experiment E909 
\cite{Igor,Tom}. This  has now been superseded by the recent
Crystal Ball data of Ref. ~\cite{koz03a,koz03b,koz03c}. In Ref.~\cite{gwunc} we obtained
one solution
very similar to that in Ref.~\cite{gw97} quoted in Table~\ref{table1}
and a second solution
$a$(fm) = 0.21+$i$0.30, $r_0$(fm) = --2.61+$i$6.67  and
$s$(fm$^3$) = --0.39--$i$3.67.
This we called an "unconventional" solution in Ref.~\cite{gwunc}, since 
not only are $a$ {\it etc} very different from before but also the form of the 
$\eta N\rightarrow\eta N$ amplitude away from the threshold is
qualitatively different. We have dwelt on this occurence of
unconventional solutions, since some of the scattering parameters in
Table~\ref{table1} could well be of this form. However, we want to
emphasize that these unconventional solutions do not seem to arise 
in the present formulation  using 
the complete data set as in A.
 
Sets E and F show the effect of removing the $\gamma N\rightarrow \eta N$ data of 
Ref.~\cite{ren02}. In this case E is a conventional solution with a 
$\chi^2$/dof= 0.946 and F an unconventional solution with an
insignificantly smaller $\chi^2$/dof= 0.942. However, this
solution has enormous  error bars, so that the corresponding $\eta N$ 
scattering parameters are very poorly determined.

Set G fits the same data as case A but now shows the effect of introducing an
energy dependence into
the basic $\pi N \rightarrow \eta N$ $K$-matrix element written in 
Eq.~(\ref{k7}). The magnitude of the dependence is governed by 
the value of $R_{\pi,\eta}$. Here we take $R_{\pi,\eta}=0.2$ fm, a value 
suggested by theory --- see the discussion after Eq.~(\ref{2}).
This is seen to have a very small effect on the values of $a, \ r_0$ and
$s$ and also on the $\chi^2$/dof, reducing the latter
 from 1.0053 for case A to 1.0052. The reason for this is clear,
since in the formalism the
$R_{\pi,\eta}$ always occurs in the combination $R_{\pi,\eta}B_{\pi,\eta}$
 and the minimization always produces a very small
value  $B_{\pi,\eta}\sim 0.01$ fm. However, the errors on $a, \ r_0$ and
$s$ are now much larger than in case A indicating that this energy
dependence is attempting to improve the  data fit far away from the
threshold. In fact, if --- as a numerical experiment --- the value of 
$R_{\pi,\eta}$ is taken as a free parameter, then it
becomes $\sim$20 fm --- an unacceptably large value. This gives scattering parameters that are similar
to set A, but away from the threshold the  
 $\eta N\rightarrow\eta N$ amplitude is qualitatively different. In
particular, the $\eta N \rightarrow \eta N$ develops a peak near 1400 MeV 
and an improvement in the fit to the $\pi N \rightarrow \pi N$ results.

So far the emphasis has been on extracting the best values for the
effective range parameters. This is certainly of interest when comparing
different approaches as in Table~\ref{table1}. However, as mentioned in
the introduction, in practical
applications involving systems with more than one nucleon, it is not the 
threshold value of the $\eta  N$ interaction that is relevant. Instead,
depending on the multinucleon system,
this interaction is needed over a range of energies upto 40 MeV below
the threshold. In Fig.~1 
the real and imaginary parts of the
$\eta  N$ interaction  are shown for sets A, B  and C in
Table~\ref{results}. There it is seen that both the real and imaginary 
parts are not only
qualitatively but also quantitatively the same at all relevant energies and
beyond.

\section{Fitting the $\eta  N\rightarrow \eta N$ amplitude with a
seperable form}
\label{sepsec}
The requirement of few-body physics is a simple separable 
approximation to the off-shell $\eta  N\rightarrow \eta N$ amplitude. 
This we provide as 
\begin{equation}
\label{toff}
T_{\eta\eta}(q, E,q')= v_{\eta}(q)t_{\eta\eta}(E) v_{\eta}(q')
\end{equation}
with $ v= 1/( 1+  q^2 \beta^2)$, where  $\beta$  is the 
range  parameter in this model, as discussed in Appendix. Now
another effective range expansion is used for $t_{\eta\eta}(E)$
\begin{equation}
t_{\eta\eta}(E)^{-1}+iq_{\eta}v(q_\eta)^2 =
1/a^s+\frac{r_0^s}{2} q^2_{\eta}+s^s q^4_{\eta}.
\end{equation}
This is to be compared with the corresponding expansion 
\begin{equation}
T_{\eta \eta}^{-1}+iq_{\eta}=1/a+\frac{r_0}{2} q^2_{\eta}+s q^4_{\eta}
\end{equation}
defined in the introduction.
The subthreshold amplitude required in few-body $\eta$-physics 
involves subthreshold energies $E$ and physical momenta $q$. 
In contrast to  $T$ the subthreshold irregularities of $t$ 
are removed in the large energy region.   

The effective range expansions of both amplitudes lead
to  the algebraic relations
\begin{equation}
\label{No6}
a^s= a ;\ r_0^s = r_0 -4\beta^2/a ; \ s^s = s - r_0 \beta^2
+3\beta^4/a.
\end{equation}
The range parameter $ \beta $ used in the separable model
is not well determined. The actual value used ($\beta=0.31$ fm) is
motivated  by the two factors  discussed after Eq.~(\ref{2}): 
the rough estimate of the $N(1535)$ formfactor 
and the distance to the nearest singularities in the $t$ and $u$ 
channels .

\section{Conclusion}
\label{conclu}
In this paper we use a $K$-matrix method, developed 
earlier~\cite{gw97,gw99}, to describe $\pi  N\rightarrow \pi  N$, $\pi 
N\rightarrow \eta N$, $\gamma N\rightarrow \pi N$ and
$\gamma N\rightarrow \eta N$ data in an energy range of about 100 MeV
or so each side of the $\eta$-threshold. Here the new feature is the
incorporation of recent $\gamma N\rightarrow \eta N$ near-threshold
 data\cite{koz03c} to replace that from the compilation of Ref.~\cite{Nefkens}.
Eventhough this new data is  not the main deciding factor for the actual values
of the $\eta$-nucleon effective range scattering parameters $a, \ r_0 $ 
and $ s$, it does play an important role in
determining the errors on these parameters. Furthermore, it appears to
make set A in Table~\ref{results} the optimal unique solution. This
is in contrast to earlier works, {\it e.g.} in Ref.~\cite{gwunc}, where
other solutions appeared with very different values of $a, \ r_{0},  \ s$ and
forms of the $\eta N$-amplitude away from threshold.
Our best and  final value of $a$= 0.91(6)+$i$0.27(2) fm is large, but since the  
$\eta N$-amplitude drops rapidly below threshold --- see
Fig.~1 
--- it is not immediately clear whether it is
sufficiently strong to develop quasi-bound states at the appropriate 
center-of-mass energies of upto to 40 MeV below the threshold. 
All that can be said is that the $\eta N$-amplitude at these
subthreshold energies is not that small as to obviously eliminate the possibility
of quasi-bound states. It could be that, in reality, the situation is
borderline and so could explain why some theories\cite{wil93,hai86,li,ued,may96}
 predict these states which, so far, are apparently not  seen 
experimentally~\cite{cal,chrien}.

 One of the authors (S.W.) wishes to acknowledge the hospitality of the
Helsinki Institute of  Physics, where part of this 
work was carried out. The authors also thank Drs.  V. V. Abaev, B. Briscoe, 
D. Rebreyend, F.  Renard and I. Strakovsky    for useful 
correspondence and private discussions concerning their  data. In
addition the authors wish to thank Dr.  N. Kelkar for pointing out
misprints in Eq.~\ref{No6}. 
This project is financed by the Academy of 
Finland contract 54038, 
 and the European Community Human 
potential Program HPRN-2002-00311 EURIDICE.

\section*{Appendix }
\label{App}
In this appendix, the physics of the above model is briefly presented.
It is based  on the nuclear resonance reaction theory of 
Wigner-Eisenbud.
The spectrum of the system consists of channel states 
$ |i)$  and an internal  single baryon state $ |o)$ of energy 
$E_o$.   
The latter may be a quark state or a bound state generated by 
some closed channels. The channel states interact via a potential $V$ 
and the propagator for such a system is denoted by  
$  g_{i,j}$ in the channel sector,  where  $  g_{o,o}= ( E-E_o)^{-1}$
and $  g_{i,o}= 0$. Coupling of the channels to the internal 
state is generated by an additional interaction $H$, and the
$ h_{i,o}= (o| H|i )$ are coupling formfactors.
The full propagator which involves both $V$ and $H$ is obtained 
from the equation 
\begin{equation}
\label{k1}
 G=  g +  g H G.  
\end{equation}
The $g$ already includes  the $V$  potential interactions and 
Eq.~(\ref{k1}) is a separable equation easy to solve. 
This gives
\begin{equation}
\label{k2}
 G_{i,j}=  g_{i,j} +   f_{i,o}G_{o,o}f_{o,j}
\end{equation}
\begin{equation}
\label{k3}
 G_{o,o}= ( E-E_o- \Sigma_o)^{-1} \      ;  \  
G_{i,o}=  g_{i,j}h_{j,o}G_ {o,o},
\end{equation}
where $\Sigma_o = h_{o,i}g_{i,i}h_{i,o} $ is the energy shift of 
the internal  state due to the channel coupling 
and $f_{i,o} = g_{i,j}h_{j,o} $. In 
these, and the following equations, the  summation over repeated 
channel indices  and integration over corresponding momenta is understood.  

The $K$-matrix is obtained in the standard way as $ K = UGU$ where $G$ 
is the standing wave propagator and, in our extended space,  $ U=V+H$. 
This results in 
\begin{equation}
\label{k4}
 K_{i,j}=  K^{pot}_{i,j} +   \gamma_{i,o}G_{o,o} \gamma_{o,j},
\end{equation}
where the potential part of the $K$-matrix is 
\begin{equation}
\label{k5}
 K^{pot}_{i,j}=  V_{i,j} +   V_{i,m}g_{m,l}V_{l,j}
\end{equation}
and the coupling formfactors become 
\begin{equation}
\label{k6}
 \gamma_{i,o}=  h_{i,o} +   V_{i,m}g_{m,j} h_{j,o}.
\end{equation}
The phenomenological model used in the main part of this work 
assumes $  \gamma_{i,o}$ to be constants   
and Eq.~(\ref{k4}) is the basis of  Eq.~(\ref{1}). 
    
In our earlier work the $K^{pot}_{i,j}$ have been considered to be 
the constants $B^{pot}_{i,j}$. Now this restriction is relaxed in the 
$\pi \eta  $ channel and   
the effective range expansion at the $\eta N $ threshold 
is made in the standard way for the potential part 
\begin{equation}
\label{k7}
 [K^{pot}(E)]^{-1}_{\pi,\eta}=  [B]^{-1}_{\pi,\eta} +  
R_{\pi,\eta}q_{\eta}^2,
\end{equation}
where  the  term $B$ refers to  the  $\eta$ threshold.
As discussed in the text there are arguments to expect a   
range term in the $\pi N$ to  $\eta N$
transition. Therefore, we invert  Eq.~(\ref{k7}) in the limited 
two channel space and obtain  
\begin{equation}
\label{k8910}
 K^{pot}_{\pi,\pi}(E) = \frac{B_{\pi,\pi} } 
{1 + 2\beta^2  q_{\eta}^2 - \gamma  q_{\eta}^4  }, \ \ \
 K^{pot}_{\eta,\eta}(E) = 
\frac{B_{\eta,\eta} } {1 + 2\beta^2  q_{\eta}^2 - \gamma  q_{\eta}^4  }, \ \ \
 K^{pot}_{\pi,\eta}(E) = \frac{B_{\pi,\eta} - D R_{\pi,\eta }q_{\eta}^2} {1 + 2\beta^2  q_{\eta}^2 - \gamma  q_{\eta}^4 },
\end{equation}
 where $ D= B_{\eta,\eta} B_{\pi,\pi}  - B_{\pi,\eta}^2 $, $ \beta^2 =B_{\pi,\eta}R_{\pi,\eta }$   
$\gamma = R_{\pi,\eta }^2  D $ 
and $  q_{\eta}^2 = ( E - E_{thr}) 2 \mu_{N \eta } $. 
For weak  potentials  parameter $\beta$ is determined directly 
by the force range. Knowing  $K_{i,j}$, the $T_{i,j}$ are given by the
usual relationship
\begin{equation}
\label{KT}
  K = \left( \begin{array}{ll}
 K_{\pi, \pi} & K_{\eta ,\pi} \\
 K_{\pi, \eta} & K_{\eta ,\eta} \end{array} \right)
 \ \ \ {\rm and} \ \
  T = \left( \begin{array}{ll}
 \frac{A_{\pi, \pi}}{1-iq_{\pi}A_{\pi, \pi}}&  \frac{A_{\eta,
 \pi}}{1-iq_{\eta}A_{\eta, \eta}}\\
  \frac{A_{\pi ,\eta}}{1-iq_{\eta}A_{\eta, \eta}}&  \frac{A_{\eta,
 \eta}}{1-iq_{\eta}A_{\eta, \eta}}\end{array} \right),
\end{equation}
where $q_{\pi,\eta}$ are the center-of-mass momenta of the two mesons in the two
channels $\pi,\eta$.
The channel scattering lengths $A_{i,j}$ are expressed in terms of the
$K$-matrix elements as
\begin{center}
$A_{\pi, \pi}=K_{\pi, \pi}+iK^2_{\pi, \eta}q_{\eta}/(1-iq_{\eta}K_{\eta ,\eta})$,
 \ \ $ A_{\eta, \pi}=K_{\eta, \pi}/(1-iq_{\pi}K_{\pi ,\pi}) $,
\end{center}
\begin{equation}
\label{2.2}
A_{\eta, \eta}=K_{\eta, \eta}+iK^2_{\eta ,\pi}q_{\pi}/(1-iq_{\pi}K_{\pi, \pi}).
\end{equation}

\begin{table}
\begin{center}
\caption{ A selection of $\eta N$- scattering lengths and effective ranges  
appearing in the literature. }
\vspace{0.5cm} 
\begin{tabular}{lcc} 
Reaction or Method & Scattering Length(fm)& Effective range (fm) \\ \hline
 \protect\cite{ben91}&0.25+$i$0.16 &\\
\protect\cite{bha85}&0.27+$i$0.22 &  \\
$pn\rightarrow d\eta$ \protect\cite{Speth}& $\leq $0.3 & \\
\protect\cite{Igor}& 0.46(9)+$i$0.18(3)&\\
 \protect\cite{mos98}& 0.487 +$i$0.171 & --6.060--$i$0.177 \\
 \protect\cite{sau95}& 0.51+$i$0.21 & \\
 \protect\cite{wil93}&0.55(20)+$i$0.30 & \\
 \protect\cite{mos98}& 0.577 +$i$0.216 & --2.807 --$i$0.057 \\
 \protect\cite{aba96}& 0.621(40)+$i$0.306(34)& \\   
 \protect\cite{kai}& 0.68+$i$0.24 & \\
\protect\cite{new}&0.717(30)+$i$0.263(25) &  \\ 
Coupled $K$-matrices\protect\cite{gw97}& 0.75(4)+$i$0.27(3)& --1.50(13) --$i$0.24(4)\\ 
$\eta d \rightarrow \eta d$ \protect\cite{Shev01}&$\geq $0.75 & \\
Coupled $K$-matrices\protect\cite{gw99}& 0.87+$i$0.27 & \\ 
\protect\cite{bat95}& 0.91(3)+$i$0.29(4)& \\
\protect\cite{ari}& 0.980+$i$0.37  & \\
 \protect\cite{pen02}& 0.991 +$i$0.347  & --2.081 --$i$0.81 \\ 
Coupled $K$-matrices\protect\cite{gw99}& 1.05+$i$0.27 & \\ 
\end{tabular}
\label{table1}
\end{center}
\end{table}
\begin{table}[h]
\begin{center} 
\caption{The $\pi N\rightarrow\eta N$ data from
Ref.~\protect\cite{koz03c}. The uncorrected $\pi$-beam momentum is
denoted by [UC] and the corrected momentum by [C].}
\begin{tabular}{ccc} \\
$P_{\pi}$ MeV/c [UC]&$P_{\pi}$ MeV/c [C]&$\sigma$ mb\\ \hline
692.5&687.1&0.64(17)\\
702.4&701.0&1.73(15)\\
714.5 &713.1&2.13(15)\\
732.1&731.6 & 2.69(20)\\
744.3 &744.1&2.68(20)\\ \hline
\end{tabular}
\label{CBdata}
\end{center}
\end{table}
\begin{table} [h]
\begin{center} 
\caption{The effective range parameters $a, \ r_0$ and $s$ using 
the notation for the elastic $T$-matrix: 
$T_{\eta \eta}^{-1}+iq_{\eta}=1/a+\frac{r_0}{2} q^2_{\eta}+s q^4_{\eta}$,
 -- with $ q_{\eta}$ being the momentum in the $\eta N$ center-of-mass.
Sets A and B fit all the data,  with A using the corrected  $\pi$-beam
momentum
in Table~\protect\ref{CBdata} and B without this correction.
Set C omits the $\pi N \rightarrow \eta N$ data of  Ref.~\protect\cite{koz03c}.
Set D is a second but unconventional solution when the data of 
Ref.~\protect\cite{koz03c} is omitted. Set E omits the 
$\gamma N \rightarrow \eta N$ data of Ref.~\protect\cite{ren02}. Set F is
a second but unconventional solution when the data of 
Ref.~\protect\cite{ren02} is omitted. 
Set G fits the same data as Set A but with an
energy dependence included in the $\pi N \rightarrow \eta N$ $K$-matrix 
element of Eq.~\protect\ref{k7}.}
\vspace{0.5cm} 
\begin{tabular}{lcccccc} \\ 
Set&Re $a$& Im $a$&Re $r_0$&Im $r_0$& Re $s$&Im $s$ \\ \hline  
A& 0.91(6)&    0.27(2)&   --1.33(15)&  --0.30(2)&   --0.15(1)&  --0.04(1)\\
B&0.88(5)&    0.25(2)&   --1.37(16)&  --0.31(2)&   --0.15(1)&   --0.04(1)\\
C&0.93(21) &   0.27(10) &  --1.3(6) & --0.31(7) &  --0.16(7)&  --0.05(3)\\
D&0.51(9)&    0.26(3)&   --2.5(6)&  0.3(5) &  0.2(2) &  --0.0(1)\\ 
E&0.77(9)&    0.25(5)&   --1.8(4)&  --0.3(1) & --0.10(69 &  --0.02(3)\\
F&0.4(5)&0.3(2)&--4(20)&2(5)&---&---\\
G&0.92(20)&0.27(9)&--1.3(6)&--0.30(6)&--0.15(6)&--0.04(3)\\
\end{tabular}
\label{results}
\end{center}
\end{table}

  \begin{figure}[ht] 
 \begin{center}
\caption{The real(R) and imaginary(I) parts of the $T(\eta$-$N)$ scattering
amplitude. The solid(dashed) lines show the fit
with(without) the pion beam correction of the $\pi N\rightarrow\eta
N$ data from Ref.~\protect\cite{koz03c} --- see
Table~\protect\ref{CBdata}.
The dotted lines show the effect of not including this $\pi N\rightarrow\eta    
N$ data in the fit. All three sets of curves are more or less indistinguishable.}
\includegraphics{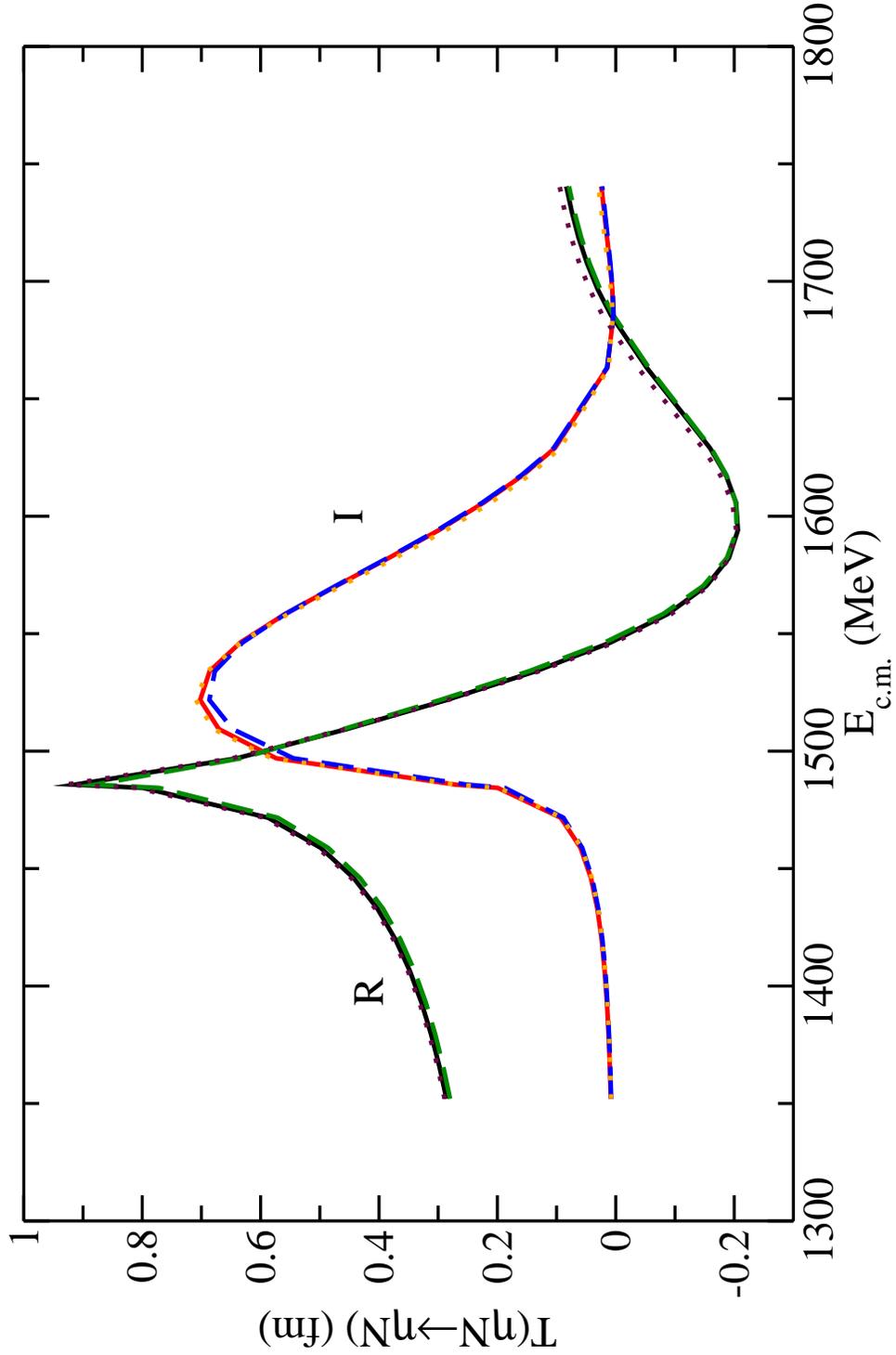}
\end{center}
\label{fig.1}
\end{figure}
\newpage 
  \begin{figure}[ht] 
 \begin{center}
\caption{Fits to the $\pi N\rightarrow\eta N$ data from 
Ref.~\protect\cite{koz03c}
with(solid line) and without(dashed line) the $\pi$-beam correction.
The squares are for the uncorrected $\pi$-beam momenta and the circles for
the corrected momenta
 --- see Table~\protect\ref{CBdata}.}
\includegraphics{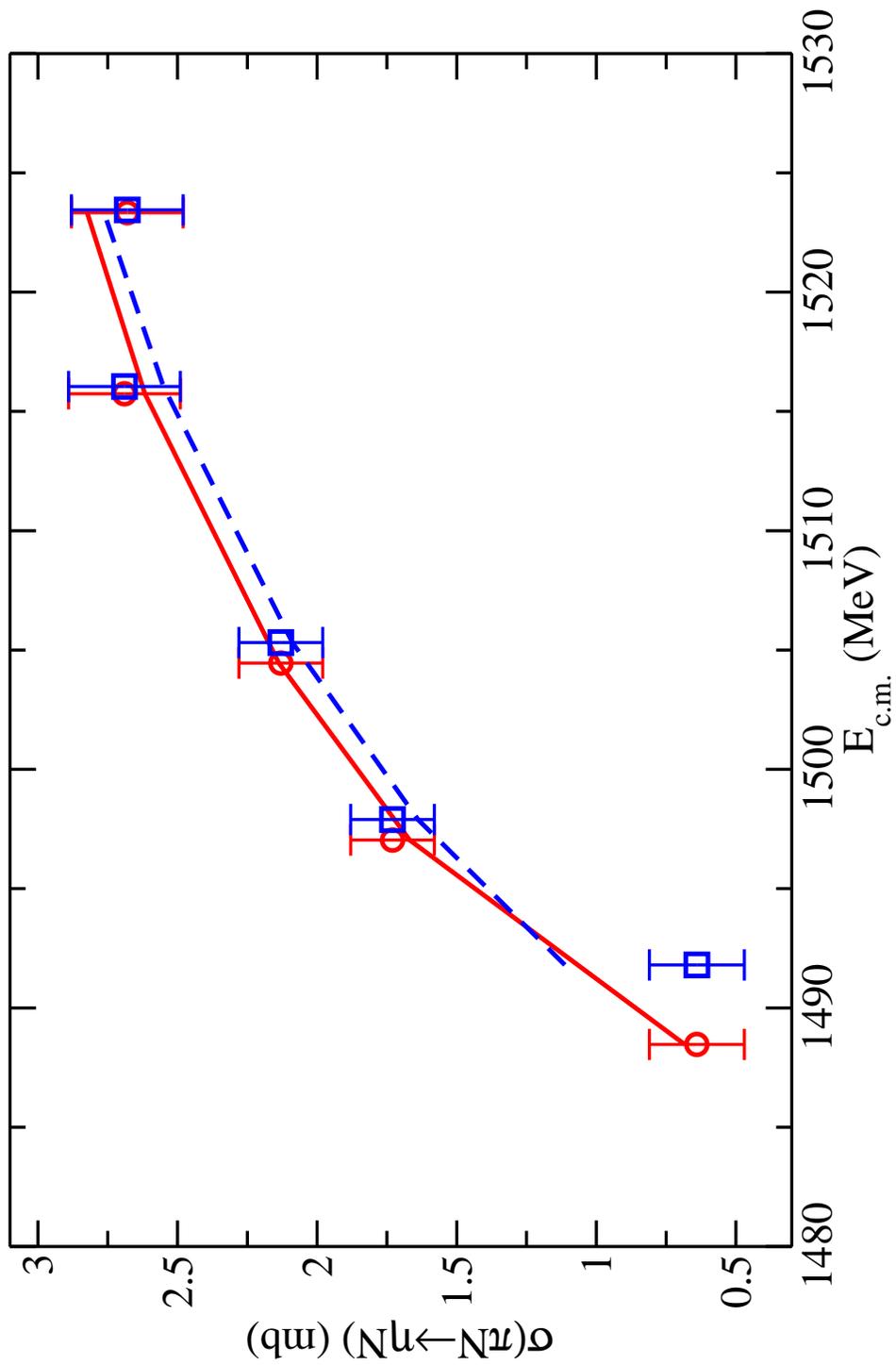}
\end{center}
\label{fig.CBfit}
\end{figure} 

\end{document}